\documentclass[fullpaper]{jpsj2}

\def\runauthor{M. Matsumoto, M. Koga, and H. Kusunose}

\title{
Single Impurity Effects in Multiband Superconductors \\
with Different Sign Order Parameters
}

\author{Masashige Matsumoto, Mikito Koga$^1$, and Hiroaki Kusunose$^2$}
\inst{
Department of Physics, Faculty of Science, Shizuoka University, 836 Ohya, Suruga-ku, Shizuoka 422--8529, Japan \\
$^1$Department of Physics, Faculty of Education, Shizuoka University, 836 Ohya, Suruga-ku, Shizuoka 422--8529, Japan \\
$^2$Department of Physics, Ehime University, Matsuyama 790-8577, Japan
}

\recdate{May 5, 2009}

\renewcommand{\H}{\mathcal{H}}
\newcommand{\br}{{\mbox{\boldmath$r$}}}

\newcommand{\bS}{{\mbox{\boldmath$S$}}}
\newcommand{\bk}{{\mbox{\boldmath$k$}}}
\newcommand{\bPsi}{{\mbox{\boldmath$\Psi$}}}
\newcommand{\bG}{{\hat{G}}}
\newcommand{\om}{{\omega_l}}
\newcommand{\omd}{{\omega^2_l}}

\newcommand{\brho}{{\hat{\rho}}}

\newcommand{\bU}{{\hat{U}}}

\newcommand{\bA}{{\hat{A}}}

\newcommand{\bskp}{{\mbox{\scriptsize\boldmath $k$}}}
\newcommand{\skp}{{\mbox{\scriptsize $k$}}}

\newcommand{\bsrp}{{\mbox{\scriptsize\boldmath $r$}}}

\newcommand{\bsk}{\bskp}
\newcommand{\sk}{\skp}

\newcommand{\bsr}{\bsrp}

\newcommand{\bbsigma}{{\mbox{\boldmath$\hat{\sigma}$}}}

\newcommand{\ri}{{\rm i}}
\newcommand{\re}{{\rm e}}
\newcommand{\rd}{{\rm d}}

\newcommand{\Tc}{{$T_{\rm c}$}}

\newcommand{\tone}{{$T_1^{-1}$}}

\newcommand{\MARU}[1]{{\ooalign{\hfil#1\/\hfil\crcr\raise.167ex\hbox{\mathhexbox20D}}}}

\newcommand{\ssp}{{$s_\pm$}}
\newcommand{\dd}{{$d_{x^2-y^2}\pm\ri d_{xy}$}}
\newcommand{\ddp}{{$d_{x^2-y^2}+\ri d_{xy}$}}

\abst{
A single impurity problem is investigated for multiband $s$-wave superconductors
with different sign order parameters (\ssp-wave superconductors) suggested in Fe-pnictide superconductors.
Not only intraband but also interband scattering is considered at the impurity.
The latter gives rise to impurity-induced local boundstates close to the impurity.
We present an exact form of the energy of the local boundstates as a function of strength of the two types of impurity scattering.
The essential role of the impurity is unchanged in finite number of impurities.
The main conclusions for a single impurity problem help us understand effects of dense impurities in the \ssp-wave superconductors.
Local density of states around the single impurity is also investigated.
We suggest impurity site nuclear magnetic resonance as a suitable experiment to probe the local boundstates that is peculiar to the \ssp-wave state.
We find that the \ssp-wave model is mapped to a chiral \dd-wave, reflecting the unconventional nature of the sign reversing order parameter.
For a quantum magnetic impurity, interband scattering destabilizes the Kondo singlet.
}

\kword{
multiband superconductivity, impurity, boundstate, Fe-pnictide superconductors, \ssp-wave superconductivity,
nuclear magnetic resonance, numerical renormalization group
}

\begin{document}

\maketitle

\section{Introduction}

The investigation of impurity effects on superconductivity has been developing for a long time.
In case of conventional $s$-wave (BCS) superconductors,
Anderson showed that nonmagnetic impurities change neither the superconducting transition temperature (\Tc)
nor the gap of the superconductor.
\cite{Anderson}
It is first pointed out by Abrikosov and Gor'kov that the magnetic impurities cause gapless behavior.
\cite{Abrikosov}
They reduce the superconducting energy gap and suppress \Tc.
As a result, they also give rise to finite density of states inside the superconducting energy gap.
\cite{Skalski,Ambegaokar,Shiba}
In the same manner as the dense magnetic impurities, a single impurity brings about localized boundstates inside the energy gap.
\cite{Soda,Fowler}
While the problem of a classical spin can be solved exactly,
\cite{Shiba,Rusinov,Sakurai}
a quantum spin involves us in a many-body problem of the Kondo effect.
\cite{Kondo}
The latter case had been studied by various theoretical methods
\cite{Matsuura,Muller-Hartmann,Jarrell}
and was finally solved for the conventional $s$-wave superconductivity using the Wilson's numerical renormalization group (NRG) method.
\cite{Wilson,Satori,Sakai}

In contrast to the BCS superconductors, nonmagnetic impurities destroy unconventional superconductivity.
For instance, Zn impurities in cupper oxide high temperature superconductors
induce additional finite density of states inside the superconducting energy gap,
which accounts for the temperature dependence of nuclear magnetic relaxation (NMR) rate.
\cite{Hotta}
A single impurity problem in $d_{x^2-y^2}$-wave superconductors was also studied.
It was found that low energy states appear with four-fold symmetry near the impurity.
\cite{Byers,Matsumoto-surface,Balatsky,Salkola,Onishi}
About a single magnetic impurity, we previously investigated a quantum spin in unconventional superconductors
using the NRG method and focused on a fully gapped chiral superconductor
expressed by $p_x\pm\ri p_y$-wave or $d_{x^2-y^2}\pm\ri d_{xy}$-wave type order parameters,
\cite{Matsumoto-Kondo-1,Matsumoto-Kondo-2,Koga-Kondo-3,Koga-Kondo-4,Fritz}
where orbital effect of the Cooper pair plays an important role.

Recently, Kamihara and coworkers discovered a new Fe-pnictide superconductors.
\cite{Kamihara,Takahashi}
It is suggested theoretically that antiferromagnetic spin-fluctuations arising from the interband nesting
favor a multiband  $s$-wave superconductivity with different sign order parameters
\cite{Kuroki,Mazin}
that is called \ssp-wave here.
For this new type of multiband superconductivity, interband scattering is important.
It affects NMR relaxation rate,
\cite{Parker,Chubukov}
can suppress \Tc,
\cite{Bang,Senga-1}
and generates impurity-induced states inside the energy gap
\cite{Senga-2,Zhang,Zhou}
similarly to a magnetic impurity in BCS superconductors.

The purpose of this paper is to investigate the single impurity problem to understand the novel properties of the \ssp-wave state.
For the single impurity, we can obtain an exact solution that helps us understand the properties of many impurity case.
The following points will be clarified in this paper:
(1) An explicit form of energy of the impurity-induced boundstates is presented
as a function of strength of the interband and intraband scatterings.
Spatial dependence of the local density of states is shown around the impurity.
(2) The pair breaking effect of the interband scattering is interpreted from an effective single band model.
Relation to chiral \dd-wave superconductors is also discussed.
(3) Quantum effect of a magnetic impurity in the \ssp-wave superconductors is analyzed by the NRG method.

This paper is organized as follows.
In \S 2, we study nonmagnetic impurity and discuss effects of interband and intraband scatterings on appearance of the localized boundstates.
In \S 3, we focus on an identical multiband case and discuss the same problem from a point of view of an effective single band model.
Then our theory is extended to a quantum magnetic impurity in \S 4.
The last section gives summary of our results.
We assume $\hbar=1$ and $k_{\rm B}=1$ throughout this paper.

\section{Nonmagnetic Impurity}

\subsection{Formulation}

Let us begin with the following Hamiltonian of a continuum model for multiband superconductivity with impurity scatterings:
\begin{align}
&\H = \sum_{\mu=\pm} \H_\mu + \H', \cr
&\H_\mu = \sum_\sigma \int\rd\br \psi_{\mu\sigma}^\dagger(\br) \epsilon_\mu(-\ri\nabla) \psi_{\mu\sigma}(\br)
      - \Delta_\mu \int\rd\br \left[ \psi_{\mu\uparrow}^\dagger(\br)\psi_{\mu\downarrow}^\dagger(\br)
                                   + \psi_{\mu\downarrow}(\br)\psi_{\mu\uparrow}(\br) \right], \cr
&\H' = \sum_{\mu\mu'=\pm}\sum_{\sigma\sigma'}
       \int\rd\br \psi_{\mu\sigma}^\dagger(\br) U_{\mu\mu',\sigma\sigma'}(\br) \psi_{\mu'\sigma'}(\br).
\label{eqn:H}
\end{align}
Here, $\H_\mu$ is the BCS Hamiltonians for the $\mu(=\pm$) band,
in which $\psi_{\mu\sigma}(\br)$ is the field operator of the conduction election of $\sigma(=\uparrow,\downarrow)$ spin
and $\epsilon_\mu(\ri\nabla)=-\nabla^2/2m_\mu-E_{\rm F}$ is the operator of the kinetic energy
for the $\mu$ band measured from the Fermi energy,
where $m_\mu$ represents the band dependent mass of the conduction electron.
$\Delta_\mu$ is the $\mu$ band superconducting order parameter.
We assume that $\Delta_\mu$ is a real value and that the sign of the order parameters are different ($\Delta_+\Delta_-<0$) between the two bands.
$\H'$ represents the Hamiltonian for the impurity scatterings.
$U_{\mu\mu',\sigma\sigma'}(\br)$ is the amplitude of the scattering
between the $\mu$ band electron with $\sigma$ spin and the $\mu'$ band electron with $\sigma'$ spin.
A single nonmagnetic impurity is located at the origin of the coordinate.
The scattering amplitude is given by
\begin{align}
U_{\mu\mu',\sigma\sigma'}(\br) = U_{\mu\mu'} \delta_{\sigma\sigma'} \delta(\br).
\label{eqn:U0}
\end{align}
Here, $\delta_{\sigma\sigma'}$ and $\delta(\br)$ are the Kronecker delta and Dirac delta functions, respectively.
The $\mu=\mu'$ components are for the intraband scattering,
while the $\mu\neq\mu'$ components are for the interband scattering.
We assume that $U_{\mu\mu'}$ is a real value and $U_{+-}=U_{-+}$.
For the nonmagnetic impurity, we define the following thermal Green's function in a $4\times 4$ matrix form:
\begin{align}
\bG(\tau,\br,\br') = - \langle T_\tau \bPsi(\br,\tau) \bPsi^\dagger(\br') \rangle,
\label{eqn:G-first}
\end{align}
where $\bPsi(\br)$ and $\bPsi^\dagger(\br)$ are 4 dimensional vectors.
The latter is defined as
\begin{align}
\bPsi^\dagger(\br) =
   \left(
     \begin{array}{cccc}
       \psi_{+\uparrow}^\dagger(\br) &  \psi_{+\downarrow}(\br) & \psi_{-\uparrow}^\dagger(\br) & \psi_{-\downarrow}(\br)
     \end{array}
   \right).
\end{align}
The imaginary-time Heisenberg representation is defined by
\begin{align}
\bPsi(\br,\tau) = \re^{\H\tau} \bPsi(\br) \re^{-\H\tau}.
\end{align}

In the absence of the impurity scattering, the unperturbed Green's function in the Fourier transformed form is given by
\begin{align}
\bG_0(\ri\om,\bk)
= \left(
    \begin{array}{cc}
      \bG_+(\ri\om,\bk) & 0 \\
      0 & \bG_-(\ri\om,\bk)
    \end{array}
  \right).
\end{align}
Here, $\om=2\pi T(l+1/2)$ is the Matsubara frequency for fermion.
$\bG_\pm$ is unperturbed $2\times 2$ Green's functions for the $\pm$ band.
It is given by
\begin{align}
\bG_\pm(\ri\om,\bk)
  = - \frac{ \ri\om + \epsilon_{\pm,\bsk}\brho_3 - \Delta_\pm\brho_1 } { \omd + \epsilon_{\pm,\bsk}^2 + \Delta_\pm^2},
\end{align}
where $\brho_\alpha~(\alpha=1,2,3)$ are the Pauli matrices for the particle-hole space.
$\epsilon_{\pm,\bsk}=k^2/2m_\pm-E_{\rm F}$ is the band dependent kinetic energy.
The real space Green's function is obtained as
\begin{align}
\bG_0(\ri\om,\br,\br') = \frac{1}{\Omega} \sum_\bsk \re^{\ri\bsk\cdot(\bsr-\bsr')} \bG_0(\ri\om,\bk).
\end{align}
Here, $\Omega$ represents the system volume.

In the presence of the impurity scattering, the Green's function is calculated exactly as
\cite{Matsumoto-surface}
\begin{align}
\bG(\ri\om,\br,\br') = \bG_0(\ri\om,\br,\br') + \bG_0(\ri\om,\br,0) \bU \left[ 1-\bG_0(\ri\om,0,0) \bU \right]^{-1} \bG_0(\ri\om,0,\br').
\label{eqn:G}
\end{align}
Here, $\bU$ is given by
\begin{align}
\bU =
\left(
  \begin{array}{cccc}
    U_{++} & 0 & U_{+-} & 0 \\
    0 & -U_{++} & 0 & -U_{+-} \\
    U_{+-} & 0 & U_{--} & 0 \\
    0 & -U_{+-} & 0 & -U_{--}
  \end{array}
\right).
\label{eqn:U}
\end{align}
$\bG_0(\ri\om,0,0)$ in eq. (\ref{eqn:G}) is calculated as
\begin{align}
\bG_0(\ri\om,0,0) &= \frac{1}{\Omega} \sum_\bsk \bG_0(\ri\om,\bk)
  =
    \left(
      \begin{array}{cc}
        \bG_+(\ri\om,0,0) & 0 \\
        0 & \bG_-(\ri\om,0,0)
      \end{array}
    \right), \cr
\bG_\pm(\ri\om,0,0) &= -\pi N_\pm \frac{\ri\om-\Delta_\pm\brho_1}{\sqrt{\omd+\Delta_\pm^2}}.
\label{eqn:G00}
\end{align}
Here, $N_\pm$ represents the density of states per volume at the Fermi energy for the $\pm$ band, respectively.
$\bG_0(\ri\om,\br,0)$ and $\bG_0(\ri\om,0,\br)$ in eq. (\ref{eqn:G}) are calculated as
\begin{align}
\bG_0(\ri\om,\br,0)=\bG_0(\ri\om,0,\br)=
    \left(
      \begin{array}{cc}
        \bG_+(\ri\om,\br,0) & 0 \\
        0 & \bG_-(\ri\om,\br,0)
      \end{array}
    \right),
\end{align}
where
\begin{align}
\bG_\pm(\ri\om,\br,0)=\frac{1}{\Omega}\sum_\bsk \re^{\ri\bsk\cdot\bsr} \bG_\pm(\ri\om,\bk).
\end{align}
For isotropic two dimensional conduction electron systems, it is expressed by Bessel functions as in eq. (\ref{eqn:Gri}).
Details of the integration are given in the Appendix.

The local density of states at position $\br$ is given by the Green's function.
Since we consider nonmagnetic scatterings here, density of states are same for $\sigma=\uparrow,\downarrow$ spins.
They are expressed as
\begin{align}
&N_\uparrow(E,\br) = N_\downarrow(E,\br) = N_{\rm intra}(E,\br) + N_{\rm inter}(E,\br), \label{eqn:dos} \\
&N_{\rm intra}(E,\br) = a_{11}(E,\br) + a_{33}(E,\br),~~~~~~N_{\rm inter}(E,\br) = a_{13}(E,\br) + a_{31}(E,\br).
\nonumber
\end{align}
Here, $N_{\rm intra}$ ($N_{\rm inter}$) is the intraband (interband) contribution.
$a_{mn}(E,\br)$ is defined by
\begin{align}
a_{mn}(E,\br) = - \frac{1}{\pi} {\rm Im} \left[ \bG(\ri\om\rightarrow E+\ri\delta,\br,\br) \right]_{mn},
\label{eqn:a_mn}
\end{align}
where $\delta$ is a positive infinitesimal small number and $[\cdots]_{mn}$ represents the $mn$ matrix element.
For dense impurities, the local density of states is uniform after averaging over the impurity positions.
In this case, only the intraband contribution remains,
since the $a_{13}$ and $a_{31}$ terms vanish after integrating over the coordinate
due to the orthogonality of the wavefunctions of different conduction bands.
For the single impurity, however, these terms remain in the presence of the interband scattering.

\subsection{Impurity-induced local boundstates}

The interband scattering connects the two bands with different sign superconducting order parameters.
We can expect boundstates as in case of many impurities.
\cite{Senga-2}
Since an important result is not altered by details of the multiband structures, we study here a case of $|\Delta_+|=|\Delta_-|=\Delta$.
In this case, we can express energy of the boundstates explicitly.
This gives us useful information about the density of states at low energy
when many impurities are taken into account.

Energy of the impurity-induced local boundstates is determined by poles of the Green's function given in eq. (\ref{eqn:G}).
Solving $|{1-\bG_0(\ri\om\rightarrow E,0,0) \bU}|=0$, we can determine the boundstate energy positions.
They are expressed explicitly as
\begin{align}
E = \pm \Delta \sqrt{
\frac{ 1 + u_{++}^2 + u_{--}^2 + \left( u_{+-}^2 - u_{++}u_{--} \right)^2 - 2u_{+-}^2 }
     { 1 + u_{++}^2 + u_{--}^2 + \left( u_{+-}^2 - u_{++}u_{--} \right)^2 + 2u_{+-}^2 }
}.
\label{eqn:bound-state-3}
\end{align}
Here, the sign of the energy ($\pm$) corresponds to particle-like and hole-like excitations, respectively.
$u_{++}$, $u_{--}$, and $u_{+-}$ are defined by
\begin{align}
u_{++} = \pi N_+ U_{++},~~~~~~
u_{--} = \pi N_- U_{--},~~~~~~
u_{+-} = \pi \sqrt{N_+ N_-} U_{+-}.
\end{align}
Let us discuss various cases below.

\subsubsection{Effect of interband scattering}

\begin{figure}[t]
\begin{center}
\includegraphics[width=6cm]{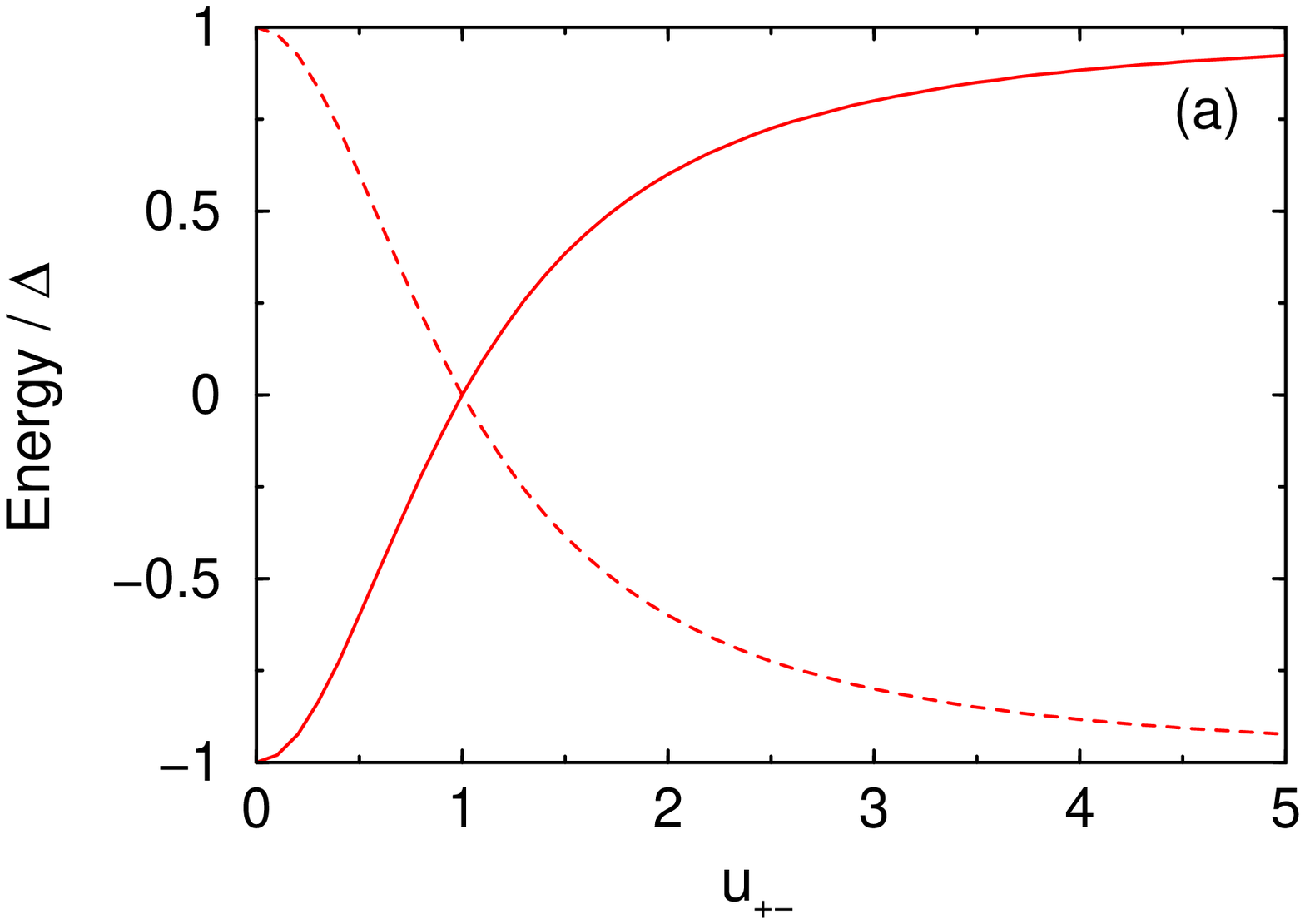}
\includegraphics[width=6cm]{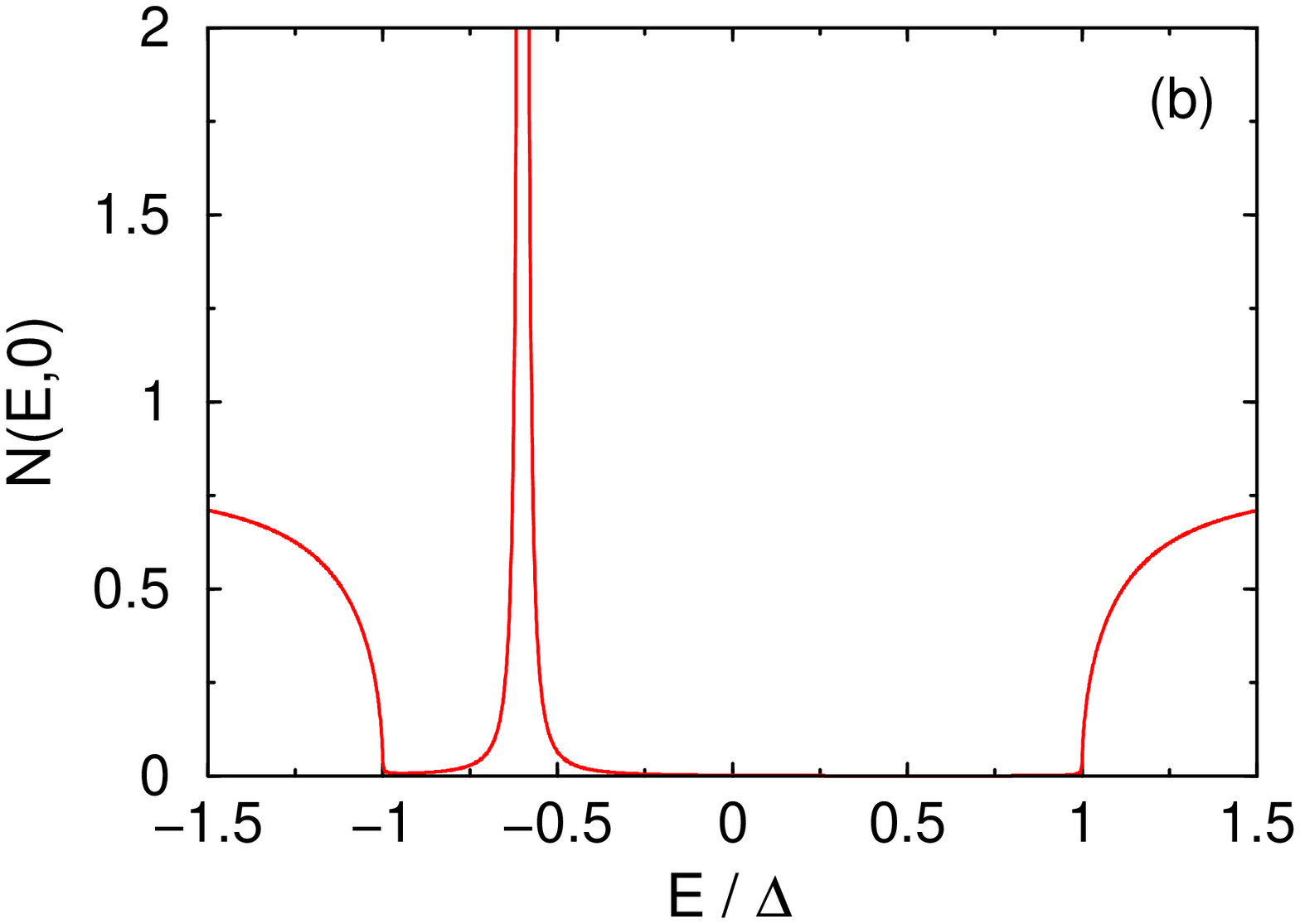}
\end{center}
\caption{
(Color online)
(a) Interband scattering dependence of the boundstate energy for $u_{++}=u_{--}=0$.
Solid and dashed lines are for $E=\pm E_{\rm B}$, respectively.
(b) Local density of states at the impurity site ($r=0$) described by eq. (\ref{eqn:dos-u12}) for $u_{+-}=0.5$.
We introduced a finite broadening factor ($\Gamma=0.001\Delta$).
}
\label{fig:bound-u12}
\end{figure}

First, we restrict ourselves to effect of the interband scattering
and discuss an identical multiband case here ($N_+=N_-$, $u_{++}=u_{--}$, and $\xi_+=\xi_-$).
When we put $u_{++}=u_{--}=0$ in eq. (\ref{eqn:bound-state-3}), we obtain
\begin{align}
E = \pm E_{\rm B},~~~~~~
E_{\rm B} = - \Delta \frac{1-u_{+-}^2}{1+u_{+-}^2} {\rm sgn}(u_{+-}).
\label{eqn:bound-state}
\end{align}
There are two boundstates corresponding to particle-like and hole-like excitations.
They intersect at $u_{+-}=1$ (or at $E=0$) as shown in Fig. \ref{fig:bound-u12}(a).
In the unitary limit ($u_{+-}\rightarrow\infty$), the boundstate energies come close to the superconducting energy gap.
This means that there is no boundstate in the unitary limit.
We notice that eq. (\ref{eqn:bound-state}) has the same form for a single classical spin in conventional $s$-wave superconductors.
\cite{Shiba,Sakurai}
In this sense, the nonmagnetic interband scattering in \ssp-wave superconductors
plays the same role of a classical spin in conventional $s$-wave superconductors.
We will discuss this point in \S 3.

Let us see the local density of states at the impurity site ($r=0$).
For the identical multiband, it is expressed as
\begin{align}
\frac{N_\sigma(E,0)}{N_++N_-} = \frac{1}{1+u_{+-}^2} \frac{ |E| \sqrt{E^2-\Delta^2}}{ E^2 - E_{\rm B}^2 } \theta(|E|-\Delta)
              + \frac{2\pi\Delta}{(1+u_{+-}^2)^2} \delta(E-E_{\rm B}),
\label{eqn:dos-u12}
\end{align}
where $\theta(x)$ is the Heviside step function.
The first term in eq. (\ref{eqn:dos-u12}) is for continuum states, while the second term is for the local boundstate.
Only one boundstate is visible as in Fig. \ref{fig:bound-u12}(b) in the identical multiband case.
Another characteristic point is that the intensity of the density of states decreases with increase of $|u_{+-}|$.
This means that the wavefunction vanishes at the impurity for large values of $|u_{+-}|$.

\begin{figure}[t]
\begin{center}
\includegraphics[width=7cm]{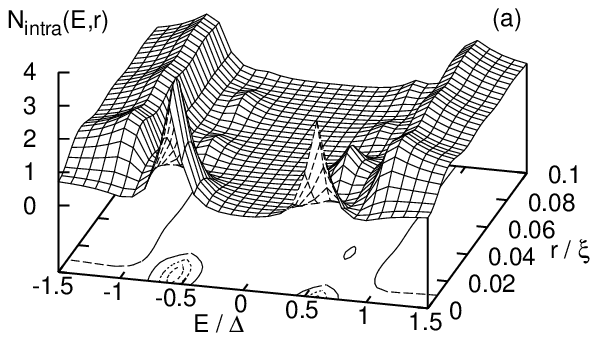}
\includegraphics[width=7cm]{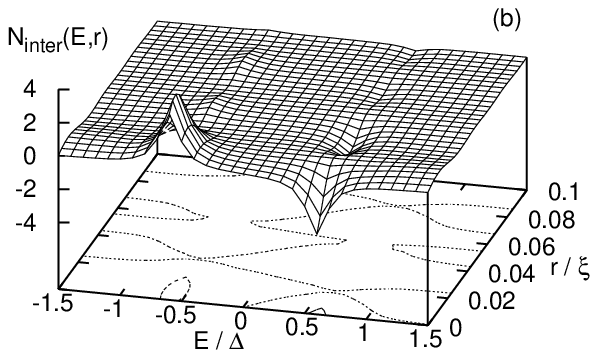}
\includegraphics[width=7cm]{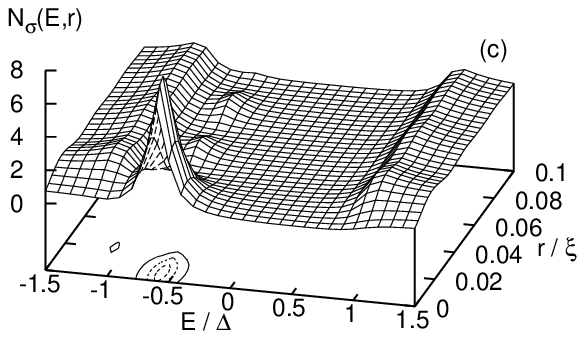}
\includegraphics[width=7cm]{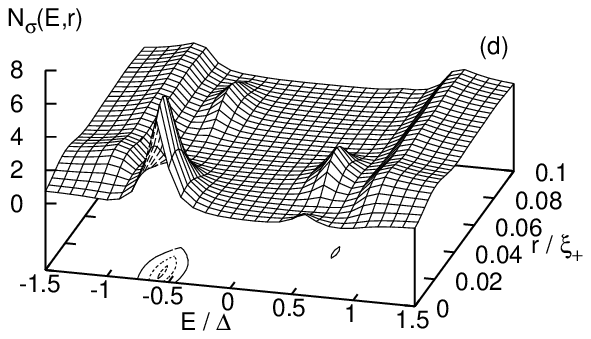}
\end{center}
\caption{
(a)-(c) Spatial dependence of the local density of states for the identical multiband
($|\Delta_+|=|\Delta_-|$, $N_+=N_-$, $u_{++}=u_{--}$, and $\xi_+=\xi_-=\xi$) renormalized by the value in the normal state.
The radius $r$ is scaled by the coherence length $\xi$ defined in eq. (\ref{eqn:xi}).
The ratio $E_{\rm F}/\Delta=100$ [see eq. (\ref{eqn:Ef-Delta})] and the broadening factor $\Gamma=0.1\Delta$ are used.
(a) $N_{\rm intra}$ for $u_{+-}=0.5$ ($u_{++}=u_{--}=0$).
(b) $N_{\rm inter}$.
(c) $N_\sigma=N_{\rm intra}+N_{\rm inter}$.
(d) $N_\sigma$ for a non-identical multiband.
Set of parameters are chosen as
$|\Delta_+|=|\Delta_-|$, $N_-=0.2N_+$, $u_{++}=u_{--}=0$, $u_{+-}=0.5$, and $\xi_+=0.2\xi_-$.
}
\label{fig:dos-x=0}
\end{figure}

We next show spatial dependence of the local density of states dividing it into the intraband [Fig. \ref{fig:dos-x=0}(a)]
and interband [Fig. \ref{fig:dos-x=0}(b)] contributions.
We can see that there are two boundstates inside the energy gap with Friedel oscillations for both continuum and boundstates.
The intensity of the boundstates decay in the superconducting coherence length $\xi$.
Adding both contributions, the density of states for the boundstate at $E=E_{\rm B}$ only remains as in Fig. \ref{fig:dos-x=0}(c)
similarly to a single classical spin in conventional $s$-wave superconductors.
\cite{Shiba,Sakurai}
This is due to the character of the identical multiband.

When the two bands are not identical, the cancelation between the two contributions is not perfect,
which results in the two boundstates inside the energy gap as in Fig. \ref{fig:dos-x=0}(d).
At the impurity site ($r=0$), the finite value for the $E=-E_{\rm B}$ boundstate is due to the $N_+\neq N_-$ character here.
Generally, $|\Delta_+|\neq|\Delta_-|$, $N_+\neq N_-$, and $u_{++}\neq u_{--}$ characters
give rise to the finite intensities for the two boundstates at $r=0$.
When those values are same for the two bands, there is no intensity for the $E=-E_{\rm B}$ boundstate at $r=0$.
The intensity at $r\neq 0$, however, can be finite when $\xi_+\neq \xi_-$,
since the cancelation becomes imperfect there.
In reality, the two bands are not identical and the appearance of particle-like and hole-like boundstates are expected inside the energy gap.
It is reported that there are two boundstates inside the energy gap in a tight-binding model calculation.
\cite{Zhang,Zhou}

\subsubsection{Effect of intraband scattering}

\begin{figure}[t]
\begin{center}
\includegraphics[width=6cm]{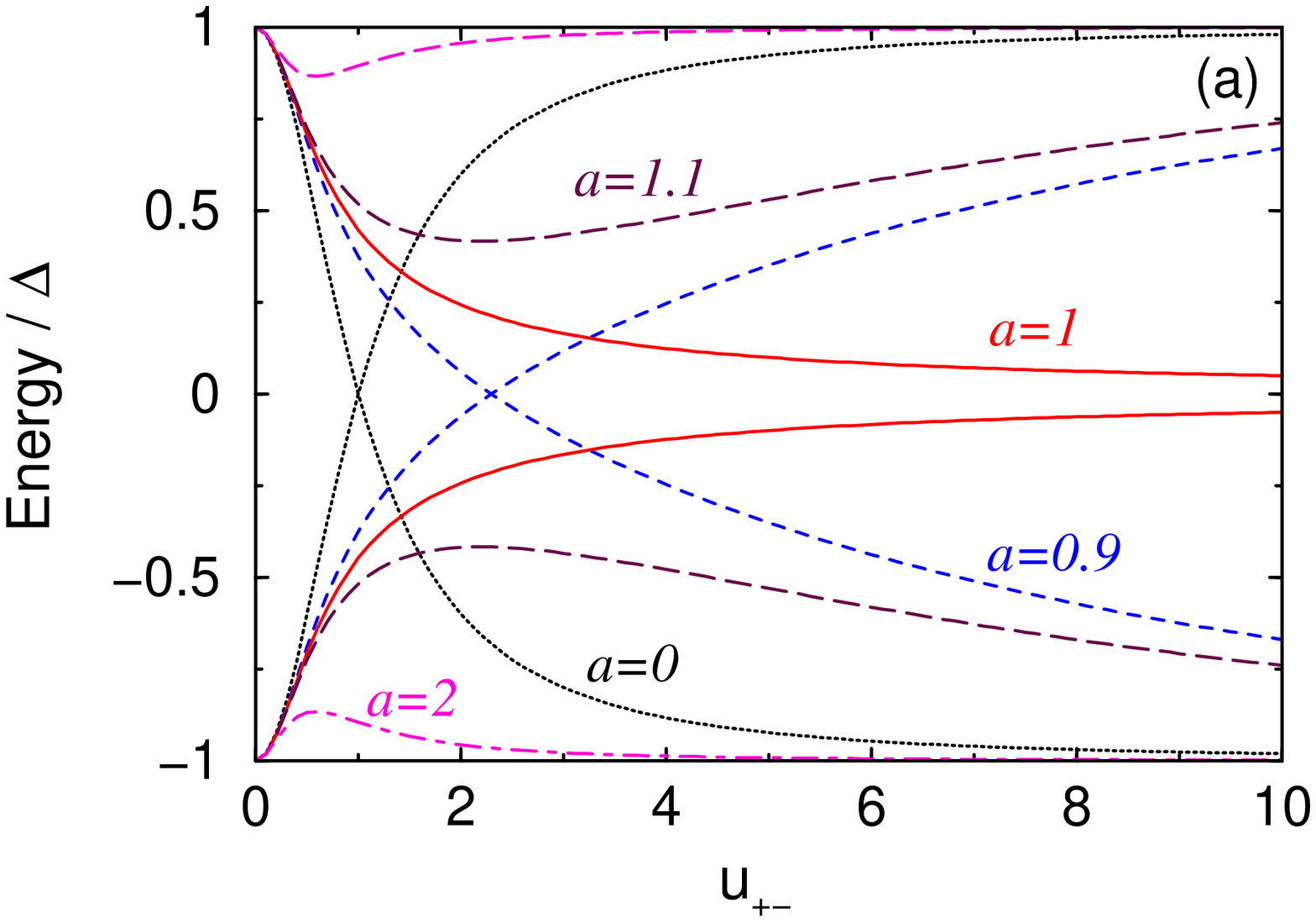}
\includegraphics[width=7cm]{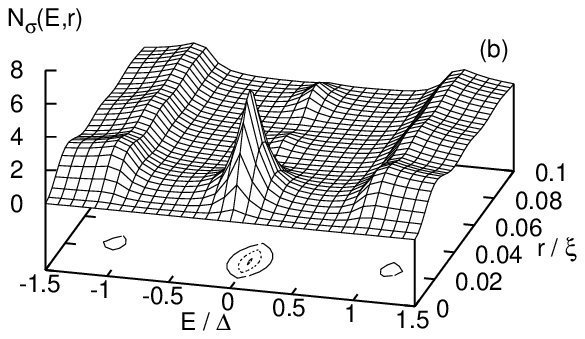}
\end{center}
\caption{
(Color online)
(a) $u_{+-}$ dependence of the boundstate energies for various $a(=u_{++}/u_{+-})$ values.
(b) Spatial dependence of the local density of states for the identical multiband in the unitary limit.
The ratio $E_{\rm F}/\Delta=100$ and the broadening factor $\Gamma=0.1\Delta$ are used.
}
\label{fig:level-x}
\end{figure}

Next, we examine effect of intraband scattering.
We consider here the identical multiband case.
The boundstate energies are given by
\begin{align}
E = \pm \Delta \frac{1+u_{++}^2-u_{+-}^2}{\sqrt{1+(u_{++}^2-u_{+-}^2)^2+2(u_{++}^2+u_{+-}^2)}}.
\label{eqn:bound-state-2}
\end{align}
We introduce a parameter $a = u_{++}/u_{+-}$ as the ratio of strength of the intraband and interband scatterings.
Figure \ref{fig:level-x}(a) shows the $u_{+-}$ dependence of the boundstate energies for various $a$.
For $a < 1$, the two boundstates intersect, while they do not for $a > 1$.
For both cases, there is no boundstate in the unitary limit.

The result is different when $a=1$ as discussed by Senga and Kontani.
\cite{Senga-1}
For $a=1$ ($u_{++}=u_{+-}\equiv u$), eq. (\ref{eqn:bound-state-2}) becomes simple as
\begin{align}
E = \pm E_{\rm B},~~~~~~
E_{\rm B} = -\Delta \frac{1}{\sqrt{1+(2u)^2}} {\rm sgn}(u).
\label{eqn:bound-state-x=1}
\end{align}
In contrast to the $a\neq 1$ case, the boundstate energy becomes $E_{\rm B}\rightarrow 0$ for $u\rightarrow\infty$.
This indicates that there is a mid-gap boundstate for $a=1$ in the unitary limit.
We note that the expression of eq. (\ref{eqn:bound-state-x=1}) is same as that
for a single nonmagnetic impurity in chiral superconductors
such as a $p_x\pm \ri p_y$-wave or $d_{x^2-y^2}\pm \ri d_{xy}$-wave type.
\cite{Okuno,Matsumoto-NMR}
For $a=1$, the impurity site local density of states is expressed as
\begin{align}
\frac{N_\sigma(E,0)}{N_+ + N_-}
=\frac{1}{1+(2u)^2} \frac{ |E| \sqrt{E^2-\Delta^2}}{ E^2 - E_{\rm B}^2 } \theta(|E|-\Delta)
+ \frac{\pi |2u|}{[1+(2u)^2]^{\frac{3}{2}}} \delta(E-E_{\rm B}).
 \label{eqn:dos0-x=1}
\end{align}
As in the $a=0$ case, there is only one boundstate ($E=E_{\rm B}$) in the local density of states due to the identical multiband character.
In Fig. \ref{fig:level-x}(b), we also show the spatial dependence of the local density of states in the unitary limit.
There is a mid-gap boundstate with Friedel oscillations.
The local density of states is much suppressed at the impurity site by the strong scattering.
We note that both boundstates ($\pm E_{\rm B}$) have finite intensities for non-identical multiband.

\subsubsection{Effect of non-identical multiband}

\begin{figure}[t]
\begin{center}
\includegraphics[width=6cm]{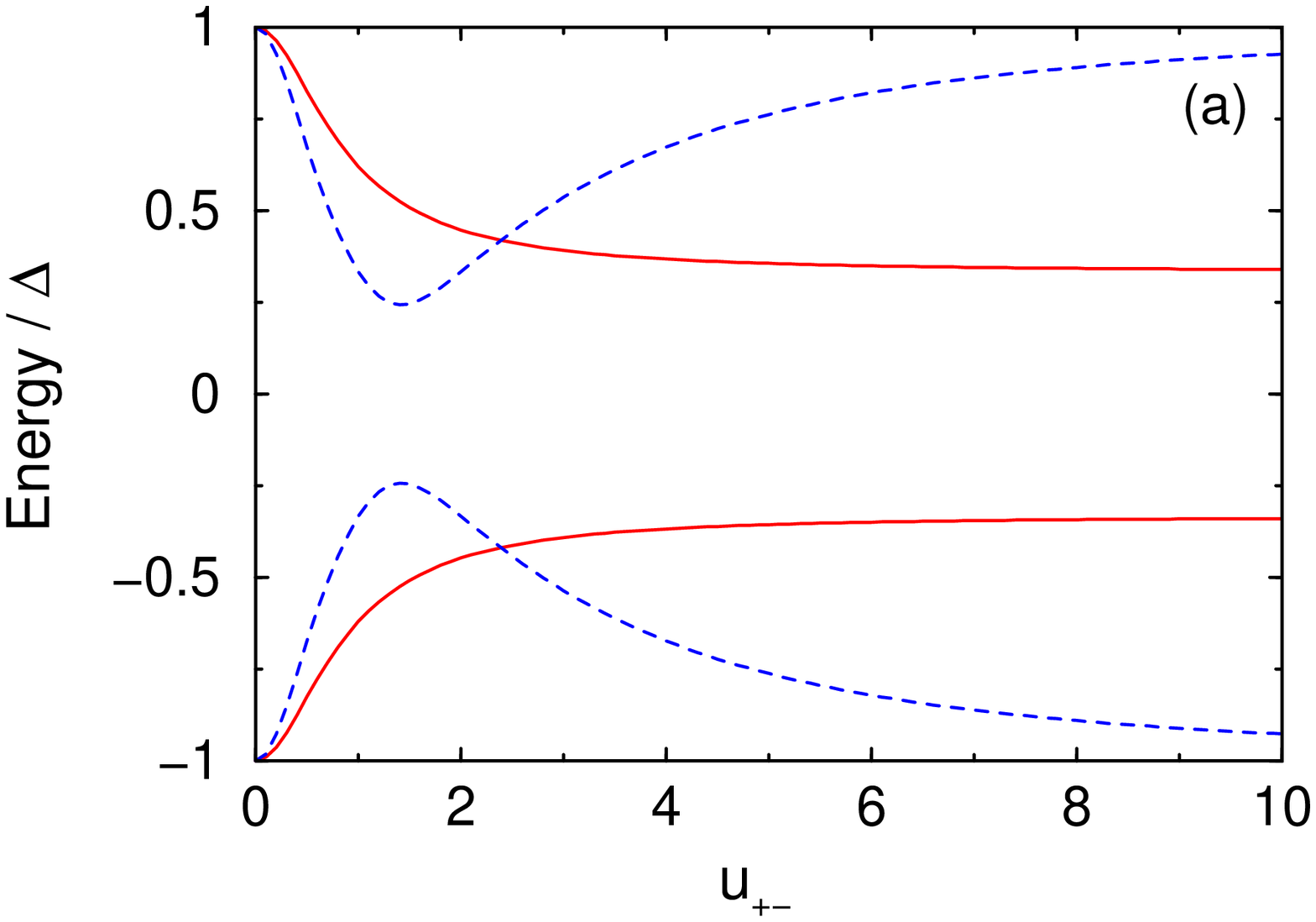}
\includegraphics[width=6cm]{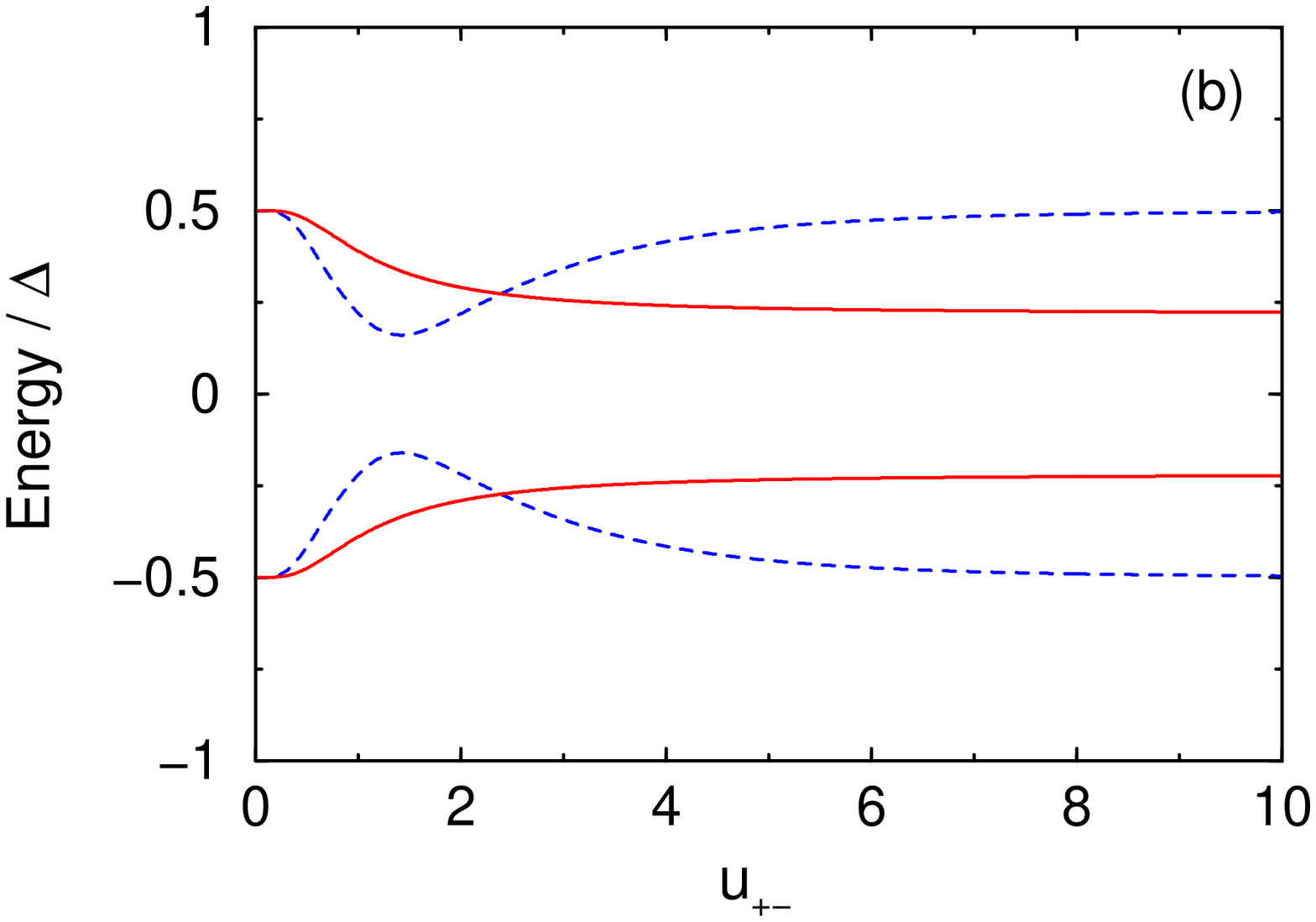}
\includegraphics[width=7cm]{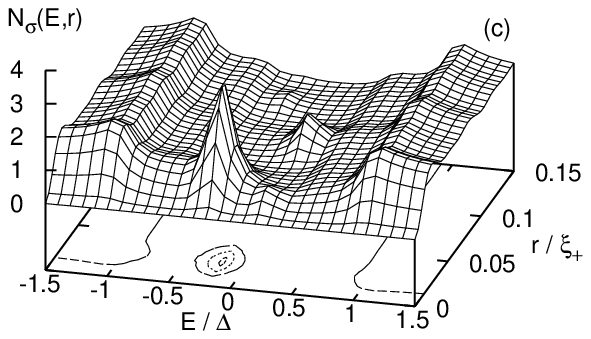}
\end{center}
\caption{
(Color online)
(a) $u_{+-}$ dependence of the boundstate energies for $|\Delta_+|=|\Delta_-|$.
The solid (dashed) line is for $u_{++}=2u_{--}=\sqrt{2}u_{+-}$ ($u_{++}=2u_{--}=u_{+-}$).
(b) For $|\Delta_-|=0.5|\Delta_+|$.
(c) Spatial dependence of the local density of states for $|\Delta_-|=0.5|\Delta_+|$ in the unitary limit.
Parameters are chosen as $u_{++}=2u_{--}=\sqrt{2}u_{+-}$ and $\xi_+=0.2\xi_-$.
The ratio $E_{\rm F}/\Delta=100$ and the broadening factor $\Gamma=0.1\Delta$ are used.
}
\label{fig:boundstate-non-identical}
\end{figure}

When the two bands are not identical, the following quantities are different:
$N_+\neq N_-$, $u_{++}\neq u_{--}$, $|\Delta_+|\neq|\Delta_-|$, $\xi_+\neq\xi_-$.
In this case, both the particle-like and hole-like boundstates have a finite intensity in the local density of states
as we discussed in the previous subsections.
Besides this, energies of the boundstates change from the identical multiband case as expressed in eq. (\ref{eqn:bound-state-3}).
In $|\Delta_+|=|\Delta_-|$ case, the boundstate energy is expressed explicitly by eq. (\ref{eqn:bound-state-3})
as a function of $u_{++}$, $u_{--}$, and $u_{+-}$.
In both the denominator and numerator, the $(u_{+-}^2-u_{++}u_{--})^2$ part is dominant in the unitary limit.
Since the boundstate energy becomes $\pm \Delta$, there is no boundstate inside the energy gap.
In contrast to this, the boundstates stay inside the energy gap even in the unitary limit
when $u_{+-}^2=u_{++}u_{--}$ (or $U_{+-}^2=U_{++}U_{--}$) is satisfied.
In this case, the boundstate energy is expressed as
\begin{align}
E = \pm \Delta \sqrt{ \frac{1+\left(u_{++}-u_{--}\right)^2}{1+\left(u_{++}+u_{--}\right)^2} }.
\end{align}
We plot the $u_{+-}$ dependence of the boundstate energy in Fig. \ref{fig:boundstate-non-identical}(a).
There are two boundstates in the unitary limit for $u_{+-}^2=u_{++}u_{--}$.

In $|\Delta_+|\neq|\Delta_-|$ case, it is difficult to express the boundstate energy explicitly.
We find poles of the Green's function and determine the boundstate energies [see Fig. \ref{fig:boundstate-non-identical}(b)].
Compared to the $|\Delta_+|=|\Delta_-|$ case, the boundstate energy shifts inside the smaller energy gap.
We also show the spatial dependence of the local density of states in Fig. \ref{fig:boundstate-non-identical}(c).
There are two boundstates inside the smaller energy gap.
Thus, the boundstate can exist even in the unitary limit when $U_{+-}^2=U_{++}U_{--}$ is satisfied.

We examined effect of a single nonmagnetic impurity with interband scattering and found that local boundstates appear near the impurity.
When there are many impurities, the local boundstates overlap each other and form an impurity band
as in the conventional $s$-wave superconductors with magnetic impurities.
\cite{Shiba}
The center of the impurity band is determined by the energy of the local boundstates,
while the width of the impurity band is controlled by impurity concentration.
Thus, eq. (\ref{eqn:bound-state-3}) is useful for knowing the in-gap state
that appears in the density of states for dense impurities as examined by Senga and Kontani.
\cite{Senga-2}

\subsection{Impurity site nuclear magnetic resonance}

For the \ssp-wave superconductivity, boundstates appear when there is an interband scattering.
Since the local boundstates exist near the impurity, we can expect those low-energy excitations to be detected by some local probes.
Scanning tunneling microscope is one of the candidates.
\cite{Pan}
Besides this, we discuss here impurity site NMR
\cite{Matsumoto-NMR}
as another candidate of a local probe to examine the exotic superconductivity of the \ssp-wave state.

The impurity site NMR relaxation rate is proportional to the following dynamical spin correlation function:
\cite{NMR-multiband}
\begin{align}
T_1^{-1} &\propto \left. \int \rd t \re^{\ri \omega t} \langle S_-(t,0) S_+(0,0) \rangle \right|_{\omega\rightarrow 0}
\propto - T \left. {\rm Im} \frac{K^{\rm R}(\omega)}{\omega} \right|_{\omega\rightarrow 0}.
\end{align}
Here, $S_\pm(t,0)$ is the spin operator in the Heisenberg representation at the impurity site ($\br=0$).
$K^{\rm R}(\omega)=K(\ri\nu_l\rightarrow\omega+\ri\delta)$ is a retarded two body Green's function.
The thermal Green's function is defined by
\begin{align}
K(\ri\nu_l) = \int_0^\beta \rd\tau \re^{\ri\nu_l\tau} K(\tau),~~~~~~
K(\tau) = - \langle T_\tau S_-(\tau,0) S_+(0,0) \rangle,
\end{align}
where $\nu_l=2\pi T l$ is a Matsubara frequency for boson.
The spin operators are written by the field operators at the impurity site.
\begin{align}
S_-(\tau,0) = \psi_{\downarrow}^\dagger(\tau,0) \psi_{\uparrow}(\tau,0),~~~~~~
S_+(0,0) = \psi_{\uparrow}^\dagger(0,0) \psi_{\downarrow}(0,0).
\end{align}
For the multiband, the field operator is written as a summation of that for the $\mu=\pm$ bands.
\cite{NMR-multiband}
\begin{align}
\psi_\sigma(0) = \sum_{\mu=\pm} \psi_{\mu\sigma}(0).
\end{align}
The two body Green's function is then written as
\begin{align}
K(\tau) &= - \sum_{\mu\mu'=\pm} \sum_{\nu\nu'=\pm}
  \langle T_\tau \psi_{\mu\downarrow}^\dagger(\tau,0) \psi_{\mu'\uparrow}(\tau,0) \psi_{\nu\uparrow}^\dagger(0,0) \psi_{\nu'\downarrow}(0,0) \rangle \cr
&= - \sum_{\mu\mu'=\pm} \sum_{\nu\nu'=\pm}
  \langle T_\tau \psi_{\mu\downarrow}^\dagger(\tau,0) \psi_{\nu'\downarrow}(0,0) \rangle
  \langle T_\tau \psi_{\mu'\uparrow}(\tau,0) \psi_{\nu\uparrow}^\dagger(0,0) \rangle \cr
&~~~+ \sum_{\mu\mu'=\pm} \sum_{\nu\nu'=\pm}
  \langle T_\tau \psi_{\mu\downarrow}^\dagger(\tau,0) \psi_{\nu\uparrow}^\dagger(0,0) \rangle
  \langle T_\tau \psi_{\mu'\uparrow}(\tau,0) \psi_{\nu'\downarrow}(0,0) \rangle.
\end{align}
Without the interband scattering, the expectation values are diagonal for the band index,
for instance,
$\langle T_\tau \psi_{\mu\downarrow}^\dagger(\tau,0) \psi_{\nu'\downarrow}(0,0) \rangle
=\langle T_\tau \psi_{\mu\downarrow}^\dagger(\tau,0) \psi_{\mu\downarrow}(0,0) \rangle \delta_{\mu\nu'}$.
In contrast to this, off-diagonal elements remain in the presence of the interband scattering.
Using the spectral representation, we can express \tone~at the impurity site as
\begin{align}
&T_1^{-1} \propto \int \rd E \frac{ g(E) \bar{g}(-E) - f(E) \bar{f}(-E) }{ 1 + \cosh(E/T) }.
\label{eqn:T1}
\end{align}
Here, $g$, $\bar{g}$, $f$, and $\bar{f}$ are expressed as
\begin{align}
&g(E) = a_{11}(E,0) + a_{33}(E,0) + a_{13}(E,0) + a_{31}(E,0), \cr
&\bar{g}(-E) = a_{22}(-E,0) + a_{44}(-E,0) + a_{24}(-E,0) + a_{42}(-E,0), \cr
&f(E) = a_{12}(E,0) + a_{34}(E,0) + a_{14}(E,0) + a_{32}(E,0), \cr
&\bar{f}(-E) = a_{21}(-E,0) + a_{43}(-E,0) + a_{23}(-E,0) + a_{41}(-E,0).
\label{eqn:gf}
\end{align}
$a_{mn}(E,\br)$ is defined by eq. (\ref{eqn:a_mn}).
In a single band case, eq. (\ref{eqn:T1}) is expressed by only the first terms in the right hand side of eq. (\ref{eqn:gf}).
In a pure (no impurity) multiband case, the second terms also remain
and the Hebel-Slichter peak is suppressed due to the cancelation of the \ssp-wave order parameters.
For the single impurity, we need the third and fourth terms (interband spin correlations) as well in the presence of the interband scattering
in the same manner as the local density of states.

\begin{figure}[t]
\begin{center}
\includegraphics[width=6cm]{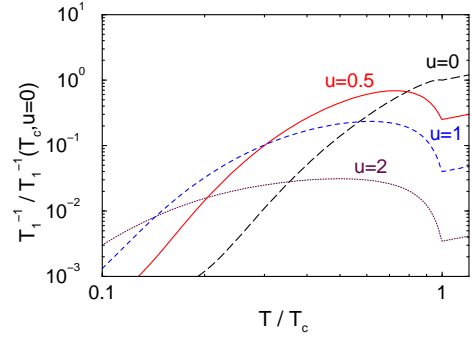}
\end{center}
\caption{
(Color online)
Temperature dependence of \tone~for the identical multiband for various values of $u$.
The scatterings are symmetric ($u_{++}=u_{--}=u_{+-}=u$).
We use the temperature dependent order parameter obtained by solving the BCS gap equation.
The broadening factor $\Gamma=0.1\Delta(0)$ is used, where $\Delta(0)$ represents the order parameter at $T=0$.
}
\label{fig:W}
\end{figure}

We show the temperature dependence of \tone~for the identical multiband in Fig. \ref{fig:W}.
For $u=0$, \tone~shows a small Hebel-Slichter peak just below \Tc~due to the canceration of the coherence factor for the \ssp-wave state.
\tone~is reduced with the increase of $u$, however, a peak appears at lower temperatures.
This does not originate from the Hebel Slichter peak but does from the impurity-induced local boundstates,
since the nuclear magnetic relaxation is possible via the local boundstates.
The temperature at the peak position is related to the energy of the boundstates.
\cite{Matsumoto-NMR}
For larger $u$, the boundstate energies decrease as in Fig. \ref{fig:level-x}(a)
and the peak position shifts towards the low temperature region as in Fig. \ref{fig:W}.
At the impurity site, impurity effects appear in the local density states strongly.
It reflects in the \tone~considerably.
Thus, the impurity site NMR is sensitive to the existence of the low energy boundstates
and it is one of a suitable probes for unconventional superconductivity.

\section{Effective Single Band Model for Identical Multiband}

In this section, we focus on the identical multiband case and discuss why the low-energy states appear
by the interband scattering on the basis of an effective single band model.
This model enables us understand the essential role of the interband scattering for the \ssp-wave state.

\subsection{Nonmagnetic scattering}

For the \ssp-wave, the Green's function and scattering matrix are defined by eqs. (\ref{eqn:G-first}) and (\ref{eqn:U})
in the $4\times 4$ matrix form, respectively.
We first diagonalize the scattering matrix.
\begin{align}
\bU^{\rm eff} = \bA^{-1} \bU \bA =
\left(
  \begin{array}{cccc}
    U_{++}+U_{+-} & 0 & 0 & 0 \\
    0 & U_{++}-U_{+-} & 0 & 0 \\
    0 & 0 & -U_{++}-U_{+-} & 0 \\
    0 & 0 & 0 & -U_{++}+U_{+-}
  \end{array}
\right),
\label{eqn:Ueff}
\end{align}
where $\bA$ is defined by
\begin{align}
\bA = \frac{1}{\sqrt{2}}
\left(
  \begin{array}{cccc}
    1 & -1 & 0 & 0 \cr
    0 & 0 & 1 & 1 \cr
    1 & 1 & 0 & 0 \cr
    0 & 0 & 1 & -1
  \end{array}
\right).
\end{align}
We next transform the Green's function by the matrix $\bA$.
\begin{align}
\bG_0^{\rm eff}(\ri\om,\bk) &= \bA^{-1} \bG_0(\ri\om,\bk) \bA \cr
&= - \frac{1}{ \omd + \epsilon_\bsk^2 + \Delta^2 }
\left(
  \begin{array}{cccc}
    \ri\om + \epsilon_\bsk & 0 & 0 & -\Delta \cr
    0 & \ri\om + \epsilon_\bsk & \Delta & 0 \cr
    0 & \Delta & \ri\om - \epsilon_\bsk & 0 \cr
    -\Delta & 0 & 0 & \ri\om - \epsilon_\bsk
  \end{array}
\right).
\label{eqn:G0-eff}
\end{align}
We notice that $\bG_0^{\rm eff}$ has the same form of a single band $s$-wave Green's function in a $4\times 4$ matrix form.
Thus, the problem reduces to a $2\times 2$ matrix form even if there is an interband scattering.
In the reduced matrix form eq. (\ref{eqn:Ueff}), we notice that $U_{+-}$ works as an Ising spin.
Therefore, the interband nonmagnetic scattering plays the role of a magnetic scattering and is pair breaking for the \ssp-wave superconductivity.

For the symmetric scattering ($U_{++}=U_{+-}$),
one of the effective potential becomes zero, while the other is $\pm 2U_{+-}$ as in eq. (\ref{eqn:Ueff}).
This means that only one of the conduction electron forming a Cooper pair is scattered by the potential, while the other is not.
Using the effective Green's function reduced to the $2\times 2$ matrix form,
we can obtain the same boundstate energy $E_{\rm B}$ defined in eq. (\ref{eqn:bound-state-2}).

\subsection{Classical magnetic scattering}

Let us consider here magnetic scattering of Ising type.
The matrix $\bU$ in eq. (\ref{eqn:U}) has the following form:
\begin{align}
\bU =
\left(
  \begin{array}{cccc}
    J_{++}^z & 0 & J_{+-}^z & 0 \\
    0 & J_{++}^z & 0 & J_{+-}^z \\
    J_{+-}^z & 0 & J_{++}^z & 0 \\
    0 & J_{+-}^z & 0 & J_{++}^z
  \end{array}
\right).
\label{eqn:U-Ising}
\end{align}
Here, $J_{++}^z$ and $J_{+-}^z$ represent coupling constants of the intraband and interband scatterings, respectively.
As in the nonmagnetic case, we obtain the following boundstate energies:
\begin{align}
E = \pm \frac{1-j_{++}^2+j_{+-}^2}{1+(j_{++}^2-j_{+-}^2)+2(j_{++}^2+j_{+-}^2)},
\label{eqn:boundstate-Ising}
\end{align}
where $j_{++}$ and $j_{+-}$ are defined by
\begin{align}
j_{++} = \pi N_+ J_{++}^z,~~~~~~
j_{+-} = \pi \sqrt{N_+ N_-} J_{+-}^z.
\end{align}
Comparing the boundstate energy for the nonmagnetic [eq. (\ref{eqn:bound-state-2})] and Ising [eq. (\ref{eqn:boundstate-Ising})] cases,
we notice that they are equivalent under the following transformations:
\begin{align}
u_{++} \longleftrightarrow j_{+-},~~~~~~
u_{+-} \longleftrightarrow j_{++}.
\end{align}
This result implies that the roles of the magnetic and nonmagnetic scatterings
are interchanged for the interband scattering in \ssp-wave superconductors.
This property has been reported by Golubov and Mazin who studied reduction of transition temperature of \ssp-wave pairing superconductors.
\cite{Golubov}

We elucidate this point by mapping the \ssp-wave model to an effective single band one.
As in the nonmagnetic case, $\bU$ is transformed as
\begin{align}
&\bU^{\rm eff} = \bA^{-1} \bU \bA =
\left(
  \begin{array}{cccc}
    J_{+-}^z+J_{++}^z & 0 & 0 & 0 \\
    0 & -J_{+-}^z+J_{++}^z & 0 & 0 \\
    0 & 0 & J_{+-}^z+J_{++}^z & 0 \\
    0 & 0 & 0 & -J_{+-}^z+J_{++}^z
  \end{array}
\right).
\end{align}
Since the $11$ and $44$ components of $\bU^{\rm eff}$ are coupled via the order parameter terms in eq. (\ref{eqn:G0-eff}),
the interband magnetic scattering $J_{+-}^z$ works as a nonmagnetic scattering
in the effective single band model in the reduced $2\times 2$ matrix form.
Thus, the roles of the magnetic and nonmagnetic scatterings are interchanged for the interband scattering in the \ssp-wave state.

\subsection{Relation between \ssp-wave and \dd-wave superconducting states}

For the \ssp-wave state, there are two superconducting conduction bands with isotropic $s$-wave order parameters.
The characteristic point is that the signs of the order parameters are opposite.
It makes difference between the \ssp-wave and the conventional $s$-wave states as we showed in the single impurity problem.
This indicates that the \ssp-wave state has unconventional pairing nature.
Since the \ssp-wave state is a fully gapped singlet pairing state,
we focus on a \dd-wave state in this subsection and discuss the single impurity problem in a different point of view.

The Hamiltonian eq. (\ref{eqn:H}) is rewritten in the momentum space.
\begin{align}
&\H = \H_{\rm kin} + \H_{\Delta} + \H_{\rm imp}, \cr
&\H_{\rm kin} = \sum_{\mu=\pm} \sum_{\bsk\sigma} \epsilon_\bsk c_{\mu\bsk\sigma}^\dagger c_{\mu\bsk\sigma}, \cr
&\H_\Delta = - \sum_{\mu=\pm} \sum_\bsk \mu \Delta \left( c_{\mu\bsk\uparrow}^\dagger c_{\mu,-\bsk\downarrow}^\dagger
                                                        + c_{\mu,-\bsk\downarrow} c_{\mu\bsk\uparrow} \right), \cr
&\H_{\rm imp} = \sum_{\mu\mu'=\pm} \sum_{\bsk\bsk'} \sum_\sigma U_{\mu\mu'} c_{\mu\bsk\sigma}^\dagger c_{\mu'\bsk'\sigma},
\label{eqn:H-NRG}
\end{align}
where $c_{\mu\bsk\sigma}^\dagger$ and $c_{\mu\bsk\sigma}$ are creation and annihilation operators for the conduction electron
with momentum $\bk$ and spin $\sigma$ for the $\mu=\pm$ band.
The Hamiltonian consists of three terms:
$\H_{\rm kin}$, $\H_\Delta$, and $\H_{\rm imp}$ are for the kinetic energy, for the pairing interaction,
and for the interaction between the conduction electron and the impurity, respectively.
Since there is a rotational symmetry around the single impurity, it is convenient to use the polar coordinate.
We transform then the operator as
\begin{align}
c_{\mu\bsk\sigma} = \sqrt{\frac{2}{\pi kR}} \sum_l (-\ri)^l \re^{\ri l\phi_\sk} c_{\mu kl\sigma},
\label{eqn:trans}
\end{align}
where $R$ represents the system size.
$l$ is the $z$ component of the orbital angular momentum of the conduction electron.
$k$ is the wave number.
$\phi_\sk$ is the angle from the wave vector measured from the $k_x$-axis.
The Hamiltonian eq. (\ref{eqn:H-NRG}) is then rewritten as
\begin{align}
&\H_{\rm kin} = \sum_{\mu=\pm} \sum_\sk \sum_\sigma \epsilon_\sk c_{\mu\sk 0\sigma}^\dagger c_{\mu\sk 0\sigma}, \cr
&\H_\Delta = - \sum_{\mu=\pm} \sum_\sk
               \mu \Delta \left( c_{\mu\sk 0\uparrow}^\dagger c_{\mu\sk 0\downarrow}^\dagger
                               + c_{\mu\sk 0\downarrow} c_{\mu\sk 0\uparrow} \right), \label{eqn:H-ss} \\
&\H_{\rm imp} = \frac{\pi k_{\rm F}R}{2} \sum_{\mu\mu'=\pm} \sum_{\sk\sk'} \sum_\sigma U_{\mu\mu'} c_{\mu\sk 0\sigma}^\dagger c_{\mu'\sk' 0\sigma}.
\nonumber
\end{align}
Here, $k_{\rm F}$ is the Fermi wave number and the summation means
$\sum_\sk = \frac{R}{\pi} \int_0^\infty \rd k.$
$\H_{\rm imp}$ is composed of the operator for the $l=0$ angular momentum.
Since the $l=0$ orbital is connected to the impurity, we retain only the $l=0$ component in the Hamiltonian.
$\H_\Delta$ represents that the total angular momentum of the Cooper pair is zero due to the $s$-wave nature of the \ssp-wave paring state.
Next, we transform the operator by
\begin{align}
c_{\mu\sk 0\sigma} = \frac{ c_{\alpha\sk\sigma} - \mu c_{\beta\sk\sigma} }{\sqrt{2}}.
\label{eqn:trans-2}
\end{align}
Here, $c_{\alpha\sk\sigma}$ and $c_{\beta\sk\sigma}$ are annihilation operators for fermion.
The Hamiltonian eq. (\ref{eqn:H-ss}) is then written as
\begin{align}
&\H_{\rm kin} = \sum_{\gamma=\alpha,\beta} \sum_\sk \sum_\sigma \epsilon_\sk c_{\gamma\sk\sigma}^\dagger c_{\gamma\sk\sigma}, \cr
&\H_\Delta = \sum_\sk \sum_\sigma \sigma
  \Delta \left( c_{\beta\sk\sigma}^\dagger c_{\alpha\sk,-\sigma}^\dagger + c_{\alpha\sk,-\sigma} c_{\beta\sk\sigma} \right), \label{eqn:H-dd2} \\
&\H_{\rm imp} = \sum_{\gamma,\gamma'=\alpha,\beta} \sum_{\sk\sk'} \sum_\sigma U_{\gamma\gamma'} c_{\gamma\sk\sigma}^\dagger c_{\gamma'\sk'\sigma}.
\nonumber
\end{align}
Here, $U_{\gamma\gamma'}$ is given by
\begin{align}
\left(
  \begin{array}{c}
    U_{\alpha\alpha} \\
    U_{\beta\beta} \\
    U_{\alpha\beta}=U_{\beta\alpha}
  \end{array}
\right)
=
\frac{1}{2}
\left(
  \begin{array}{c}
     U_{++} + U_{--} + 2U_{+-} \\
     U_{++} + U_{--} - 2U_{+-} \\
    -U_{++} + U_{--}
  \end{array}
\right).
\label{eqn:UU}
\end{align}

To discuss physical meaning of eq. (\ref{eqn:H-dd2}), we consider Hamiltonian $\H_\Delta$ for the \ddp-wave pairing state.
\begin{align}
\H_\Delta = - \sum_\bsk \Delta_\bsk \left( c_{\bsk\uparrow}^\dagger c_{-\bsk\downarrow}^\dagger + c_{-\bsk\downarrow} c_{\bsk\uparrow} \right).
\label{eqn:H-dd}
\end{align}
Here, $\Delta_\bsk = \Delta \re^{2\ri\phi_\sk}$ is the momentum dependent order parameter for the \ddp-wave state.
Substituting eq. (\ref{eqn:trans}) into eq. (\ref{eqn:H-dd}), we obtain
\cite{Koga-Kondo-4}
\begin{align}
\H_\Delta = \sum_\sk \sum_\sigma \sigma
 \Delta \left( c_{2\sk\sigma}^\dagger c_{0\sk,-\sigma}^\dagger + c_{0\sk,-\sigma} c_{2\sk\sigma} \right), \label{eqn:H-dd3}
\end{align}
where irrelevant angular momentum components disconnected to the impurity are truncated here.
The total angular momentum of the Cooper pair is expressed by $+2$, reflecting the chiral \ddp-wave character.
We notice that eq. (\ref{eqn:H-dd3}) has the same form of $\H_\Delta$ in eq. (\ref{eqn:H-dd2}).
Therefore, the $\alpha$ and $\beta$ indices introduced in eq. (\ref{eqn:trans-2})
correspond to the $l=0$ and $l=2$ angular momentum in the \ddp-wave picture.
Thus, the single impurity problem in the \ssp-wave state is equivalent to that in the \dd-wave state.

Let us consider the symmetric scattering case ($U_{++}=U_{--}=U_{+-}=U$) in the \ddp-wave picture.
Since only $U_{00}=2U$ is finite [$U_{\alpha\alpha}=2U$ in eq. (\ref{eqn:UU})],
it can be mapped to a short-range scattering impurity problem in the \dd-wave state.
It is known that the short-range scattering gives rise to a local boundstate of energy given in eq. (\ref{eqn:bound-state-x=1}).
\cite{Okuno,Matsumoto-NMR}
It explains why the value of $2u$ appears in eq. (\ref{eqn:bound-state-x=1}).
Therefore, it is natural to have the mid-gap state in the unitary limit for the symmetric scattering in \ssp-wave superconductors.

\begin{figure}[t]
\begin{center}
\includegraphics[width=7cm]{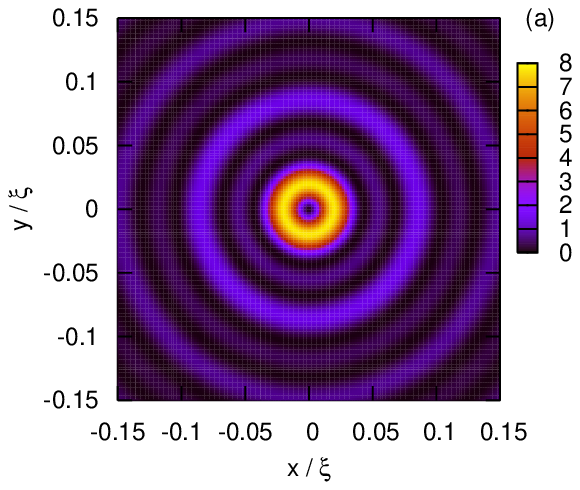}
\includegraphics[width=7cm]{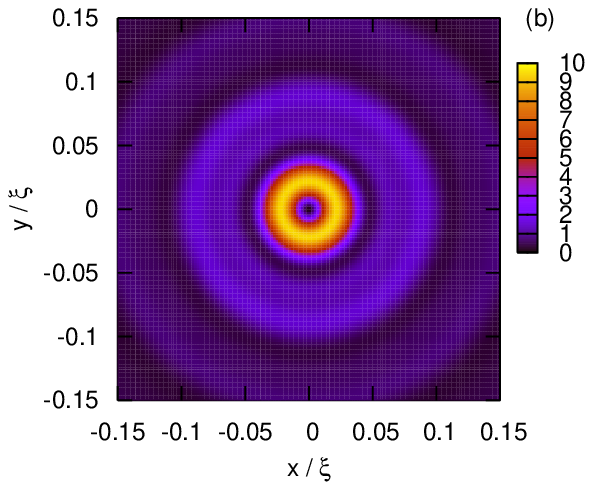}
\end{center}
\caption{
(Color online)
Local density of states at $E=0$ (mid-gap) in the unitary limit.
(a) \ssp-wave case with the symmetric scattering.
(b) For \dd-wave.
The impurity is located at the origin of the coordinate.
The radius $r$ is scaled by the coherence length $\xi$.
The ratio $E_{\rm F}/\Delta=100$ and the broadening factor $\Gamma=0.08\Delta$ are used.
}
\label{fig:dos-impurity}
\end{figure}

Although energy of the boundstate is same in the \ssp-wave and \dd-wave states,
a little difference between them appears in spatial dependence of the local boundstates
around the impurity as shown in Fig. \ref{fig:dos-impurity},
since the $\alpha$ ($\beta$) index introduced in eq. (\ref{eqn:trans-2}) is not the angular momentum $l=0$ ($l=2$).
The real space Green's function for the \ddp-wave is given in the Appendix.
The difference between the two cases can be seen only in a microscopic length scale (Fermi wave length).
In a long length scale such as the superconducting coherence length, there is no significant difference between them.

We mention here another different point between the \ssp-wave and \dd-wave states.
Since the \dd-wave state breaks the time reversal symmetry,
electric current is induced by scatterings such as an impurity, surface, and domain wall.
\cite{Okuno-2,Matsumoto-Ru}
In contrast to this, the time reversal symmetry is not broken in the \ssp-wave state and such current is not induced.

\section{Quantum Spin and Kondo Effect}

It is known that magnetic impurities destroy the superconducting order parameter
and suppress the superconducting transition temperature.
Although these results are for conventional BCS superconductors,
they hold also in the \ssp-wave superconductors when the magnetic scattering is intraband type.
However, the interband type is open to further investigation.
In this section, we examine effects of the interband magnetic scattering in \ssp-wave superconductors in the identical multiband case
using the Wilson's NRG method
\cite{Wilson}
which is reliable to study the Kondo effect also in superconductors.
\cite{Satori,Sakai}

In the same manner as the nonmagnetic scattering [$\H_{\rm imp}$ in eq. (\ref{eqn:H-ss})],
the Hamiltonian for the magnetic impurity is expressed as
\begin{align}
&\H_{\rm imp} = \frac{\pi k_{\rm F}R}{2} \sum_{\mu\mu'=\pm} \sum_{\sk\sk'} \sum_{\sigma\sigma'} \bS\cdot\bbsigma_{\sigma\sigma'}
  J_{\mu\mu'} c_{\mu\sk 0\sigma}^\dagger c_{\mu'\sk' 0\sigma'},
\label{eqn:H-NRG-mag}
\end{align}
where $\bS$ represents the $S=1/2$ spin operator for the impurity.
$J_{++}$ and $J_{--}$ ($J_{+-}=J_{-+}$) are for the intraband (interband) magnetic scattering.

\begin{figure}[t]
\begin{center}
\includegraphics[width=6cm]{fig7a.eps}
\hspace{1cm}
\includegraphics[width=3cm]{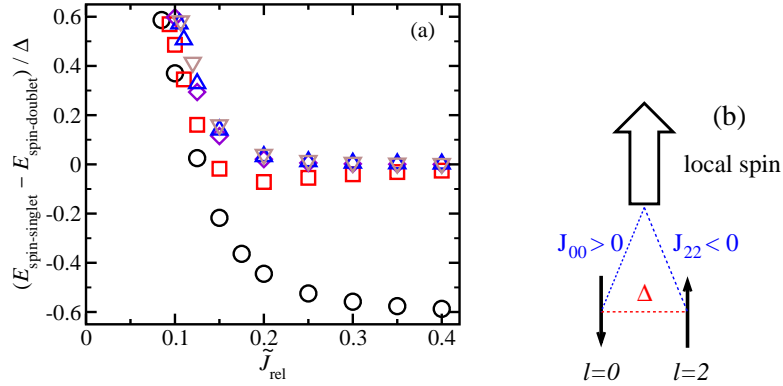}
\end{center}
\caption{
(Color online)
(a) The coupling $J_{\mu\mu'}$ dependence of the boundstate energy levels for $S=1/2$ local spin in the \ssp-wave state.
The data are for $b=J_{+-}/J_{++}$ ($J_{++}=J_{--}$),
the ratio of strength of the interband and intraband scatterings:
$b = 0$ (circle), 0.1 (square), 0.5 (diamond), 1.0 (up-triangle) and 2.0 (down-triangle).
$J_{\rm rel} = (J_{++} + J_{+-})$ is the relevant coupling in the Kondo effect (see text)
and $\tilde{J}_{\rm rel} = 1.5 N_0 J_{\rm rel}$ is used for our NRG analysis.
(b) Schematic of the groundstate in the \ddp-wave picture.
$J_{00}$ is antiferromagnetic, while $J_{22}$ is ferromagnetic.
$\Delta$ represents the Cooper pairing interaction between the $\downarrow$ ($l=0$) and $\uparrow$ ($l=2$) conduction electrons.
}
\label{fig:NRG}
\end{figure}

We examine the magnetic impurity problem as in the \ddp-wave case.
\cite{Koga-Kondo-4}
In Fig. \ref{fig:NRG}(a), for various values of
$b = J_{+-} / J_{++}$,
we show the energy of the lowest-lying spin-singlet state with particle-hole degeneracy
measured from that of the lowest-lying spin-doublet at low temperatures in respect of the relevant coupling
$J_{\rm rel} = (J_{++} + J_{+-})$, where $J_{++} = J_{--}$ and $N_+=N_-\equiv N_0$ are assumed.
The meaning of $J_{\rm rel}$ is described later.
Although appearance of the boundstates can be understood qualitatively by the Ising spin case in \S 3.2, for the quantum spin,
there are two energy scales characterizing the competition of superconducting pairing and Kondo-singlet formation.
One is the superconducting energy gap $\Delta$ and the other is the Kondo temperature
defined simply as $T_{\rm K} = N_0 J_{++} \exp (-1 / N_0 J_{++})$ for $b = 0$ in the unit of the half width of band.
For $b = 0$, the ground state changes from the spin-doublet state to the spin-singlet as $J_{++}$ increases.
The Kondo singlet is stabilized for a large $T_{\rm K} / \Delta$ only when $J_{+-}=0$.
This resembles the case of a local $S = 1/2$ quantum spin in a conventional $s$-wave superconductor.
Once $J_{+-}$ is finite, the spin-singlet energy merges into the spin-doublet one for a large $J_{++}$.
This implies that $J_{+-}$ destabilizes the Kondo singlet.
Besides $b \simeq 0$, the qualitative behavior is represented by the $b = 1$ case described by
an $s$-wave scattering magnetic impurity coupled to the chiral \dd-wave superconductivity discussed below.
On the other hand, for a small $T_{\rm K} / \Delta$, the doubly degenerate bound (spin-singlet) state
appears in the superconducting energy gap, like a nonmagnetic impurity in \S 2.
One can also find that $T_{\rm K}$ is estimated to be $N_0 J_{\rm rel} \exp (-N_0 J_{\rm rel})$.

Let us discuss the above result for the \ssp-wave state in terms of the \ddp-wave picture.
$\H_{\rm imp}$ can be mapped to the \ddp-wave model as in the nonmagnetic scattering given in eq. (\ref{eqn:H-dd2}).
There is the following relation in the coupling constants:
\begin{align}
\left(
  \begin{array}{c}
    J_{00} \\
    J_{22} \\
    J_{20}=J_{02}
  \end{array}
\right)
= J_{++}
\left(
  \begin{array}{c}
     1 + b \\
     1 - b \\
     0
  \end{array}
\right).
\label{eqn:JJ}
\end{align}
Let us consider here that the scattering is only the interband type ($J_{++}=J_{--}\rightarrow 0$).
In this case ($b\gg 1$), the coupling constants are $J_{00}=-J_{22}$ in the \ddp-wave model [see eq. (\ref{eqn:JJ})].
This set of parameter means that one of the $J_{00}$ and $J_{22}$ is antiferromagnetic and the other is ferromagnetic.
Figure \ref{fig:NRG}(b) is a schematic of the groundstate in the \ddp-wave picture.
We can see that there is no frustration and the spin doublet ground state is stabilized even for a strong coupling,
since $J_{22}<0$ is ferromagnetic here.
This picture still holds for $J_{++} = J_{--} \ne 0$ as long as $J_{+-}$ is finite.
This means that the larger one of $J_{00}$ or $J_{22}$ is relevant.
Therefore $J_{\rm rel}$ discussed above is equivalent to $J_{00}$ that stabilizes the Kondo singlet only with one orbital.

\section{Summary}

In this paper, we investigated single impurity effects in \ssp-wave superconductors.
The main results of this paper are as follows:

(1) Energy of the impurity-induced local boundstate is expressed explicitly
as a function of strength of nonmagnetic interband and intraband scatterings [see eq. (\ref{eqn:bound-state-3})].
The result for the single impurity problem is related to the energy level of the in-gap state
that appears in the density of states for many impurities.
\cite{Senga-2}
Spatial dependence of the local density of states provides information for a local probe such as scanning tunneling microscope.

(2) We suggested impurity site NMR experiment as a powerful probe of the local boundstates induced by the nonmagnetic interband scattering.
It would capture some features of the \ssp-wave superconductivity.

(3) Roles of the magnetic and nonmagnetic interband scatterings are interchanged in the \ssp-wave superconductors.
We elucidated this point by mapping the \ssp-wave multiband model to an effective $s$-wave single band one.

(4) Appearance of the single-impurity-induced local boundstates in \ssp-wave superconductors can be understood
by a single impurity model in chiral fully gapped \dd-wave superconductors.
The \ssp-wave pairing state has similar unconventional nature of the \dd-wave superconductivity.

(5) For a quantum magnetic impurity case,
we found that the interband scattering destabilizes the Kondo singlet with two band electrons.
Appearance of the boundstates can be understood by a classical spin qualitatively,
while the boundstate energy depends on $T_{\rm K} / \Delta$.

\acknowledgements

This work is supported by a Grant-in-Aid for Scientific Research (No. 20540353) for the Japan Society for the Promotion of Science.
One of the authors (H. K.) is supported by a Grant-in-Aid for Scientific Research on Innovative Areas ``Heavy Electrons'' (No. 20102008)
of The Ministry of Education, Culture, Sports, Science, and Technology, Japan.

\appendix
\section{Real space Green's function}

\subsection{\ssp-wave state}

In this appendix, we calculate the real space Green's function.
We assume isotropic two dimensional conduction electron system.
For the $\mu=\pm$ band, the Green's function is given by
\begin{align}
\bG_\mu(\ri\om,\br,0) &= - \frac{1}{\Omega} \sum_\bsk
                           \frac{\ri\om + \epsilon_{\mu\bsk}\brho_3-\Delta_\mu\brho_1}{\omd+\epsilon_{\mu\bsk}^2+\Delta_\mu^2}
                       \re^{\ri\bsk\cdot\bsr} \cr
&\simeq - \frac{N_\mu}{2\pi} \int_{-\pi}^{\pi} \rd \phi_\sk \int_{-\infty}^\infty \rd \epsilon
   \frac{\ri\om + \epsilon\brho_3-\Delta_\mu\brho_1}{\omd+\epsilon^2+\Delta_\mu^2}
   \re^{\ri\sqrt{k_{\mu\rm F}^2+2m_\mu\epsilon}r\cos\phi_\sk}.
\label{eqn:Gr}
\end{align}
Here, $r$ is the radius from the center position of the impurity.
$\phi_\sk$ is the angle of the wave vector measured from the $k_x$-axis.
We divide eq. (\ref{eqn:Gr}) into two parts.
The first is proportional to $\ri\om-\Delta_\mu\brho_1$, and the second is proportional to $\epsilon\brho_3$.
We perform the integral of these parts independently.
The first is calculated as
\begin{align}
I_1 &= \int_{-\pi}^{\pi} \rd \phi_\sk \int_{-\infty}^\infty \rd \epsilon
       \frac{1}{\omd+\epsilon^2+\Delta_\mu^2} \re^{\ri\sqrt{k_{\mu\rm F}^2+2m_\mu\epsilon}r\cos\phi_\sk} \cr
    &= \frac{-\ri}{2\Omega_{\mu l}} \int_{-\pi}^{\pi} \rd \phi_\sk \int_{-\infty}^\infty \rd \epsilon
       \left( \frac{1}{\epsilon-\ri\Omega_{\mu l}} - \frac{1}{\epsilon+\ri\Omega_{\mu l}} \right)
       \re^{\ri\sqrt{k_{\mu\rm F}^2+2m_\mu\epsilon}r\cos\phi_\sk} \cr
    &= \frac{2\pi}{2\Omega_{\mu l}} \int_{-\pi/2}^{\pi/2} \rd \phi_\sk
         \left( \re^{\ri\sqrt{k_{\mu\rm F}^2+2m_\mu\ri\Omega_{\mu l}}r\cos\phi_\sk}
              + \re^{-\ri\sqrt{k_{\mu\rm F}^2-2m_\mu\ri\Omega_{\mu l}}r\cos\phi_\sk} \right) \label{eqn:I1} \\
    &= \frac{2\pi^2}{2\Omega_{\mu l}} \left[ J_0(k_{\mu+}r) + \ri H_0(k_{\mu+}r) + J_0(k_{\mu-}r) - \ri H_0(k_{\mu-}r) \right],
\nonumber
\end{align}
where $J_0(z)$ and $H_0(z)$ are the 0th Bessel and Struve functions, respectively.
They are defined by
\begin{align}
J_0(z) = \frac{1}{\pi} \int_{-\pi/2}^{\pi/2} \rd \phi_\sk \cos(z\cos\phi_\sk),~~~~~~
H_0(z) = \frac{1}{\pi} \int_{-\pi/2}^{\pi/2} \rd \phi_\sk \sin(z\cos\phi_\sk).
\end{align}
In eq. (\ref{eqn:I1}), we used 
\begin{align}
\Omega_{\mu l} = \sqrt{\omd+\Delta_\mu^2},~~~~~~
k_{\mu\pm} = \sqrt{k_{\mu\rm F}^2\pm 2m_\mu\ri\Omega_{\mu l}}.
\label{eqn:Omega}
\end{align}
The integral of the second part is calculated as
\begin{align}
I_2 &= \int_{-\pi}^{\pi} \rd \phi_\sk \int_{-\infty}^\infty \rd \epsilon
       \frac{\epsilon}{\omd+\epsilon^2+\Delta_\mu^2} \re^{\ri\sqrt{k_{\mu\rm F}^2+2m_\mu\epsilon}r\cos\phi_\sk} \cr
    &= \frac{1}{2} \int_{-\pi}^{\pi} \rd \phi_\sk \int_{-\infty}^\infty \rd \epsilon
       \left( \frac{1}{\epsilon-\ri\Omega_{\mu l}} + \frac{1}{\epsilon+\ri\Omega_{\mu l}} \right)
       \re^{\ri\sqrt{k_{\mu\rm F}^2+2m_\mu\epsilon}r\cos\phi_\sk} \cr
    &= \frac{\ri 2\pi}{2} \int_{-\pi/2}^{\pi/2} \rd \phi_\sk
         \left( \re^{\ri\sqrt{k_{\mu\rm F}^2+2m_\mu\ri\Omega_{\mu l}}r\cos\phi_\sk}
              - \re^{-\ri\sqrt{k_{\mu\rm F}^2-2m_\mu\ri\Omega_{\mu l}}r\cos\phi_\sk} \right) \label{eqn:I2} \\
    &= \ri\frac{2\pi^2}{2} \left[ J_0(k_{\mu+}r) + \ri H_0(k_{\mu+}r) - J_0(k_{\mu-}r) + \ri H_0(k_{\mu-}r) \right].
\nonumber
\end{align}
Using $I_1$ and $I_2$, we obtain the real space Green's function as
\begin{align}
\bG_\mu(\ri\om,\br,0) &= \pi N_\mu \frac{-\ri\om+\Delta_\mu\brho_1}{2\Omega_{\mu l}}
                       \left[ J_0(k_{\mu+}r) + \ri H_0(k_{\mu+}r) + J_0(k_{\mu-}r) - \ri H_0(k_{\mu-}r) \right] \cr
&~~~+ \ri\pi N_\mu \frac{-\brho_3}{2} \left[ J_0(k_{\mu+}r) + \ri H_0(k_{\mu+}r) - J_0(k_{\mu-}r) + \ri H_0(k_{\mu-}r) \right] \label{eqn:Gri} \\
&= - \pi N_\mu \frac{1}{2\Omega_{\mu l}} \left\{
    \left[ J_+(k_{\mu+}r) + J_-(k_{\mu-}r) \right] \left( \ri\om - \Delta_\mu\brho_1 \right)
  + \left[ J_+(k_{\mu+}r) - J_-(k_{\mu-}r) \right] \ri\Omega_{\mu l} \brho_3 \right\},
\nonumber
\end{align}
where $J_\pm(z)$ are defined by
\begin{align}
J_\pm(z) = J_0(z)\pm\ri H_0(z).
\label{eqn:Jpm}
\end{align}
Equation (\ref{eqn:Gri}) reduces to eq. (\ref{eqn:G00}) for $r=0$, since $J_0(0)=1$ and $H_0(0)=0$.
In the practical calculation, we perform the integrals in eqs. (\ref{eqn:I1}) and (\ref{eqn:I2}) numerically.
It is convenient to introduce the following band dependent coherence length $\xi_\mu$ and dimensionless radius $\bar{r}_\mu$:
\begin{align}
\xi_\mu = \frac{v_{\mu\rm F}}{2\Delta_\mu} = \frac{k_{\mu\rm F}}{2m_\mu}\frac{1}{\Delta_\mu},~~~~~~
r = \xi_\mu \bar{r}_\mu.
\label{eqn:xi}
\end{align}
We can rewrite $k_{\mu\pm}r$ in eq. (\ref{eqn:I1}) as
\begin{align}
k_{\mu\pm}r = \sqrt{1+\ri\frac{\Omega_{\mu l}}{E_{\rm F}}} \frac{E_{\rm F}}{\Delta_\mu} \bar{r}_\mu.
\label{eqn:Ef-Delta}
\end{align}

\subsection{\dd-wave state}

For the \ddp-wave state, the order parameter depends on the wavevector ($\Delta_\bsk = \Delta \re^{\ri 2\phi_\sk}$).
Green's function in a $2\times 2$ matrix form is given by
\begin{align}
\bG_0(\ri\om,\br,0) &= - \frac{1}{\Omega} \sum_\bsk \frac{\ri\om + \epsilon_\bsk\brho_3-\Delta\re^{\ri 2\phi_\sk} \brho_1}
                                                         {\omd+\epsilon_\bsk^2+\Delta^2} \re^{\ri\bsk\cdot\bsr} \cr
&\simeq - \frac{N_0}{2\pi} \int_{-\pi}^{\pi} \rd \phi_\sk \int_{-\infty}^\infty \rd \epsilon
   \frac{\ri\om + \epsilon\brho_3-\Delta\re^{\ri 2\phi_\sk}\brho_1}{\omd+\epsilon^2+\Delta^2}
   \re^{\ri\sqrt{k_{\rm F}^2+2m\epsilon}r\cos\phi_\sk}.
\end{align}
The term proportional to $\Delta$ is different form the \ssp-wave case.
This term is calculated as
\begin{align}
I_3 &= \int_{-\pi}^{\pi} \rd \phi_\sk \int_{-\infty}^\infty \rd \epsilon
       \frac{\Delta\re^{\ri 2\phi_\sk}}{\om^2+\epsilon^2+\Delta_\mu^2} \re^{\ri\sqrt{k_{\rm F}^2+2m\epsilon}r\cos\phi_\sk} \cr
    &= \frac{2\pi\Delta}{2\Omega_l} \int_{-\pi/2}^{\pi/2} \rd \phi_\sk \re^{\ri 2\phi_\sk}
       \left( \re^{\ri\sqrt{k_{\rm F}^2+2m\ri\Omega_l}r\cos\phi_\sk}
            + \re^{-\ri\sqrt{k_{\rm F}^2-2m\ri\Omega_l}r\cos\phi_\sk} \right) \label{eqn:I3} \\
    &= \frac{2\pi^2\Delta}{2\Omega_l} \left\{ -J_2(k_+r) + \ri \left[ H_0(k_+ r) - \frac{2H_1(k_+r)}{k_+r} \right]
                                              -J_2(k_-r) - \ri \left[ H_0(k_- r) - \frac{2H_1(k_-r)}{k_-r} \right] \right\},
\nonumber
\end{align}
where $J_2(z)$ and $H_1(z)$ are the second Bessel and the first Struve functions, respectively.
In eq. (\ref{eqn:I3}), we used 
\begin{align}
\Omega_l = \sqrt{\omd+\Delta^2},~~~~~~
k_\pm = \sqrt{k_{\rm F}^2\pm 2m\ri\Omega_l}.
\end{align}
Using $I_1$, $I_2$, and $I_3$, we obtain the real space Green's function for the \dd-wave.
In the same manner as eq. (\ref{eqn:xi}), we introduce a coherence length for the \dd-wave.
We perform the integral in eq. (\ref{eqn:I3}) numerically as in eqs. (\ref{eqn:I1}) and (\ref{eqn:I2}).


\end{document}